\documentclass{pss}
\usepackage{graphicx}
\usepackage{amsmath}

\begin{document}

\title{Metal Nanowires: Quantum Transport, Cohesion, and Stability}
\author{C. A. Stafford\thanks{e-mail: 
stafford@physics.arizona.edu; fax: (520) 621-4721}
}
\address{Physics Department, University of Arizona, 1118 East 4th St., 
Tucson, AZ 85721}
\submitted{\today}

\maketitle


\begin{abstract}
Metal nanowires exhibit a number of interesting properties: their
electrical conductance is quantized, their shot-noise is suppressed by the
Pauli principle,
and they are remarkably strong and stable.
We show that many of these properties can be understood
quantitatively using a nanoscale generalization of the free-electron model.
Possible technological applications of nanowires are also discussed.
\end{abstract}

\section{Introduction}

Metal nanowires represent nature's ultimate limit of conductors down to
a single atom in thickness.
In the past eight years, experimental
research on 
metal nanowires has burgeoned 
\nocite{Olesen,Krans,Rubio96,Durig,Kondo97,liquid,Scheer,Auchain,Auchain2,Brom99,Yanson,Kondo00,Ugarte01} [1-13].
The simplest model of a metal is the free-electron model \cite{Ashcroft76},
which already describes many bulk properties of simple monovalent metals
semiquantitatively.  In this article, we discuss our generalization of 
the free-electron model to describe nanoscale conductors 
\nocite{StaffordPRL97,Kassubek99,BuerkiPRB99,BuerkiPRL99,StaffordPRL99,StaffordPRL00,StaffordEDP00,Kassubek01} [15-22].

\begin{figure} 
\resizebox{9cm}{!}{
\includegraphics*[11cm,4cm][200mm,85mm]{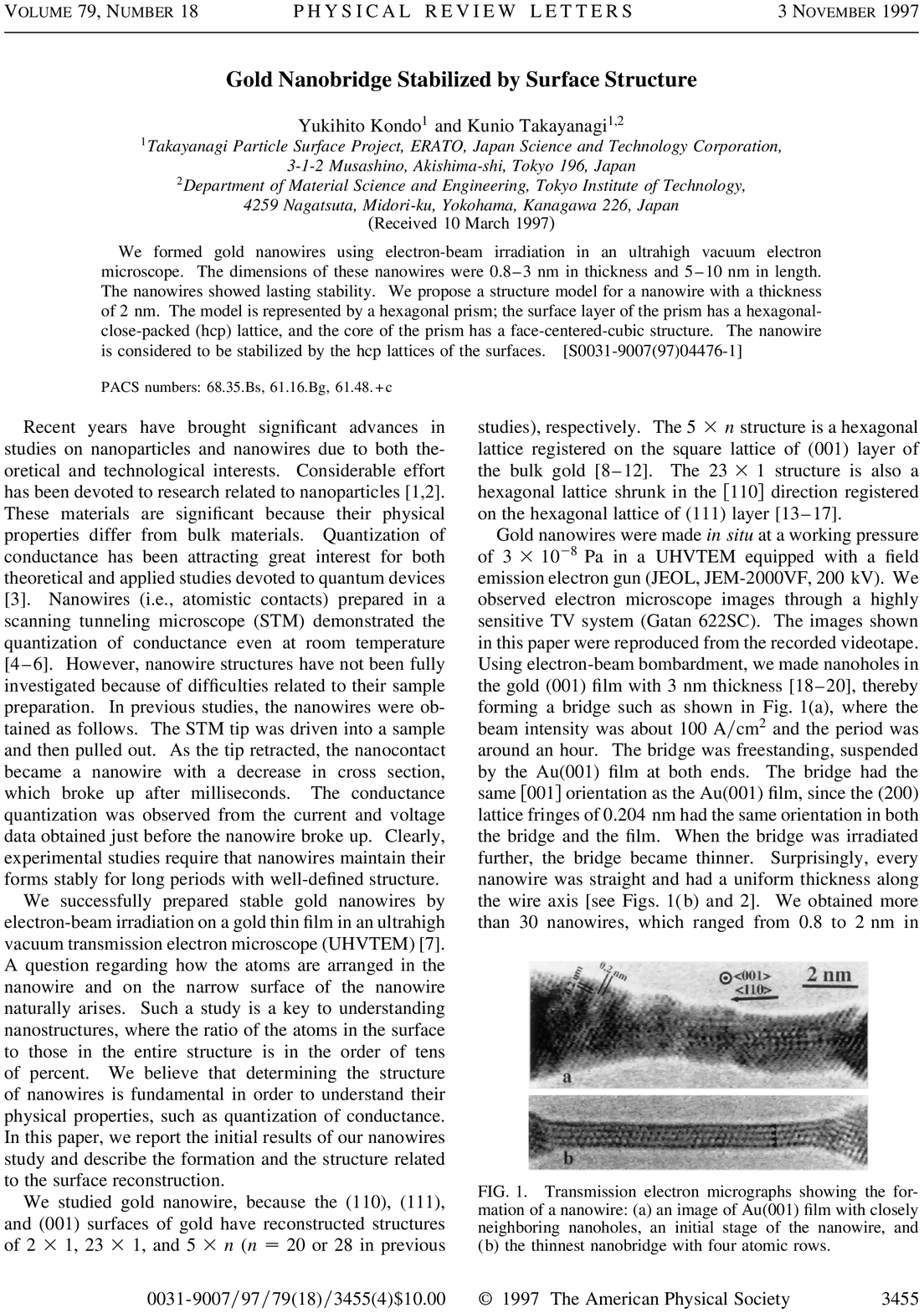}}
\caption{Transmission electron micrographs showing the formation of a
gold nanowire \cite{Kondo97} (image courtesy of Y.\ Kondo): 
({\bf a}) an image of Au(001)
film with closely spaced nanoholes, the initial stage of the nanowire;
({\bf b}) a nanowire four atoms in diameter, resulting from further
electron-beam irradiation.}
\label{fig:nanobridge}
\end{figure}

A remarkable feature of metal nanowires is the fact that they are stable at
all.  Fig.\ \ref{fig:nanobridge} 
shows electron micrographs by Kondo and Takayanagi
\cite{Kondo97} illustrating the formation
of a gold nanowire.  Under electron beam irradiation, the wire becomes
ever thinner, until it is but four atoms in diameter.  Almost all of the
atoms are at the surface, with small coordination numbers. The surface energy
of such a structure is enormous, yet it is observed to form spontaneously, and
to persist almost indefinitely.  Even wires one atom thick
are found to be remarkably stable 
\cite{Auchain,Auchain2,Ugarte01}.  
Naively, such structures might be expected to
break apart into clusters due to surface tension \cite{Chandrasekhar},
but we find that electron-shell effects can stabilize arbitrarily
long nanowires \cite{Kassubek01}.

\begin{figure} 
\resizebox{5.5cm}{!}{
\includegraphics*[11cm,54mm][20cm,18cm]{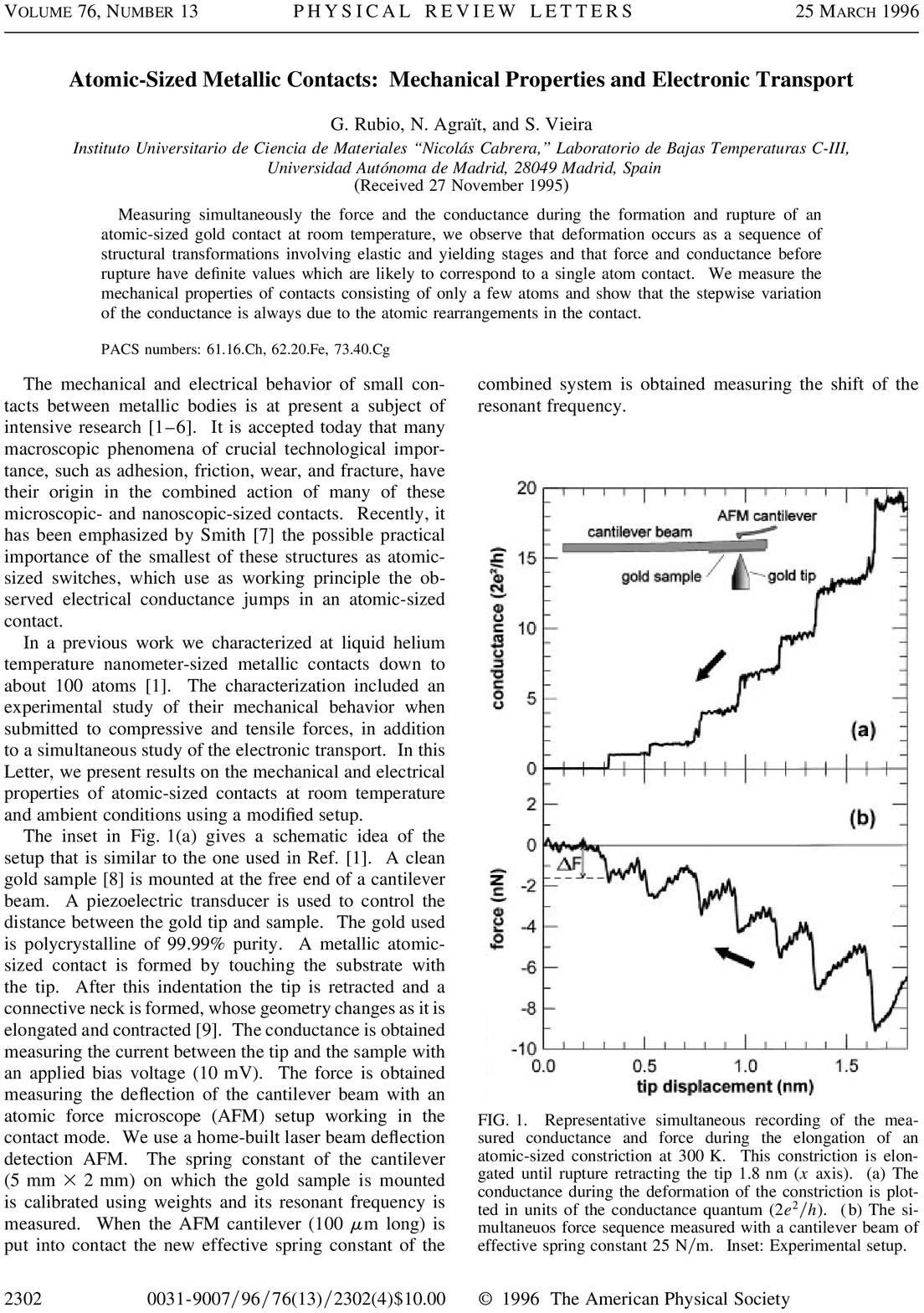}}
\resizebox{7.5cm}{!}{\includegraphics{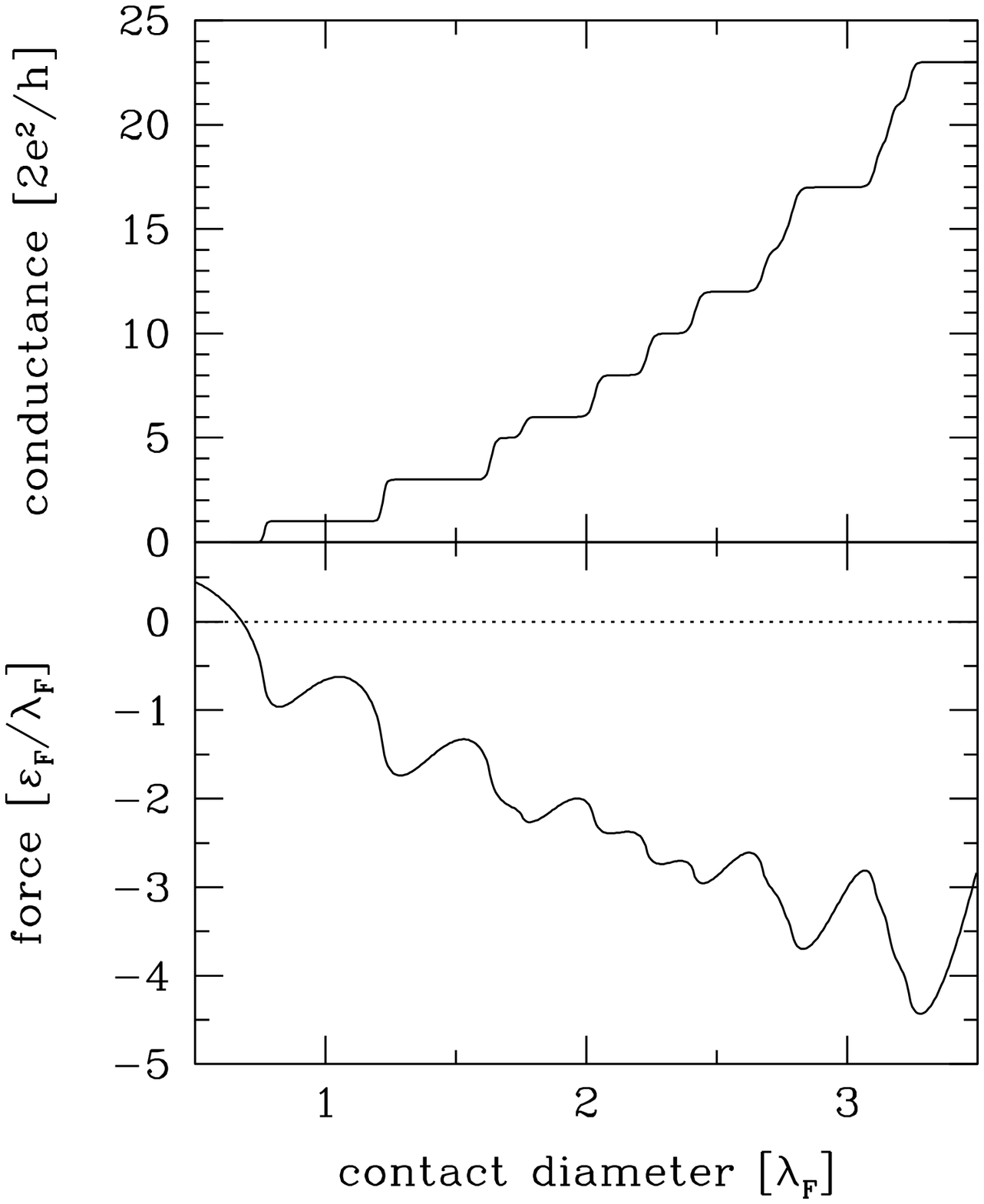}}
\caption{Left: Measured electrical conductance and cohesive
force of a gold nanowire \cite{Rubio96} (image courtesy
of N.\ Agra\"{\i}t).  Right: Calculated conductance 
and force of a metal nanowire, modeled as a
constriction in a free-electron gas with hard walls
\cite{StaffordPRL97}. 
Note that $\varepsilon_F/\lambda_F\simeq 1.7\mbox{nN}$ in gold.}
\label{fig:fandg}
\end{figure}

A crucial clue to understanding the physics of metal nanowires 
is the observed correlation 
between their electrical and mechanical properties.
In a seminal experiment \cite{Rubio96}
carried out in 1995, Rubio, Agra\"{\i}t and Vieira
simultaneously measured the electrical conductance and cohesive force of
an atomic-scale gold wire as it formed and ruptured
(see Fig.\ \ref{fig:fandg}, left panel).
They observed steps of order $G_0=2e^2/h$ in the conductance, which were
synchronized with a sawtooth structure with an amplitude
of order 1nN in the force.  Similar results were obtained independently by
Stalder and D\"urig \cite{Durig}. 
Note that the tensile strength of the nanowire
in the final stages before rupture exceeds
that of macroscopic gold by a factor of 20,
and is of the same order of magnitude as 
the theoretical value in the absence of dislocations \cite{Rubio96}.
This is consistent with the recent finding of Rodrigues, Fuhrer, and Ugarte 
that such nanowires are, in fact, typically free of defects in their central
region \cite{Ugarte01}.

The standard description of nanoscale cohesion, pioneered by Landman and
coworkers \cite{Landman90},
is via molecular dynamics simulations \cite{Landman90,Todorov93,Brandbyge97},
which utilize 
short-ranged interatomic potentials suitable to describe the bulk
properties of metals.  However, 
such an approach appears problematic
when applied to metal nanowires, in which 
electron-shell effects \cite{Yanson} due to the transverse
confinement are likely to be important.
On the other hand, atomistic quantum calculations \cite{lda}
using, e.g., the local-density approximation,
are restricted to such small systems that their results
can not really be disentangled from finite-size effects \cite{StaffordPRL00}.
An alternative approach, developed by our group,
is to replace the discrete
ionic coordinates by a coarse-grained jellium background, in order
to be able to treat the electronic degrees of freedom correctly.
We have argued \cite{StaffordPRL97} that an atomic-scale contact between two
pieces of metal can be thought of as a waveguide for conduction electrons
(which are responsible for both electrical conduction and cohesion
in simple metals): Each quantized mode transmitted through the contact
contributes $2e^2/h$ to its conductance and a force
of order $\varepsilon_F/\lambda_F$ (roughly 1nN) to its cohesion,
where $\lambda_F$ is the de Broglie
wavelength of an electron at the Fermi energy $\varepsilon_F$ (see Fig.\
\ref{fig:fandg}, right panel).  To my knowledge, our approach is 
the only one in which the observed correlations between the
cohesive and conducting properties of metal nanowires have been explained
within a single theoretical model.

The paper is organized as follows:  The free-electron model of nanoscale
conductors is introduced in the next section, followed by a discussion of 
quantum transport, including the effect of realistic contacts to the nanowire.
Nanoscale cohesion is then analyzed within our model, followed by a 
discussion of the remarkable stability of nanowires.  The paper concludes with
some comments about the technological promise of metal nanowires.

\section{Free-electron model}

We investigate the simplest possible model
\cite{StaffordPRL97,Kassubek99} for a metal
nanowire: a free (conduction) electron gas confined within the wire by
Dirichlet boundary conditions.  A nanowire is an open quantum system, and
so is treated most naturally in terms of the electronic scattering matrix
$S$.  The Landauer formula \cite{Datta} 
expressing the electrical conductance in terms of the submatrix
$S_{12}$ describing transmission through the wire is
\begin{equation}
G=\frac{2e^2}{h} \int d\varepsilon 
\frac{-\partial f(\varepsilon)}{\partial \varepsilon} \mbox{Tr}\left\{
S_{12}^\dagger(\varepsilon) S_{12}(\varepsilon)\right\} 
\stackrel{T\rightarrow 0}{=}
\frac{2e^2}{h} \sum_{n} T_{n}(\varepsilon_F),
\label{eq:landauer}
\end{equation}
where $f(\varepsilon)$ is the Fermi-Dirac distribution function and the 
transmission probabilities $\{T_n\}$ are
the eigenvalues of $S_{12}^\dagger S_{12}$.
The conductance of a metal nanocontact was calculated exactly in this model
by Torres {\em et al.} \cite{Torres}.
The appropriate thermodynamic potential to describe the energetics of
such an open system is the grand canonical potential $\Omega$: 
\begin{equation}
\Omega = -\frac{1}{\beta} \int d\varepsilon\, 
g(\varepsilon) \ln \left(1+e^{-\beta(\varepsilon-\mu)}\right)
\stackrel{T\rightarrow 0}{=} 
\int_0^{\varepsilon_F} d\varepsilon \, g(\varepsilon) 
(\varepsilon - \varepsilon_F),
\label{eq:omega_def}
\end{equation}
where $\beta$ is the inverse temperature, $\mu$ is the
chemical potential of electrons injected into the nanowire from the
macroscopic electrodes, and $g(\varepsilon)$ is the
electronic density of states (DOS) of the nanowire.  The DOS of an open
system may be expressed in terms of the scattering matrix as
\cite{Dashen}
\begin{equation}
g(\varepsilon) = 
\frac{1}{2\pi i} \mbox{Tr} \left\{S^{\dagger}(\varepsilon)\frac{\partial S}{
\partial \varepsilon} - \mbox{H.c.}
\right\}.
\label{eq:dos_dmb}
\end{equation}
This formula is also known as the Wigner delay.  Thus, once the electronic
scattering problem for the nanowire is solved, both 
transport and energetic quantities
can be readily calculated \cite{StaffordPRL97,Kassubek99,BuerkiPRB99}.
Electron-electron interactions can be included at the mean-field
level in this model in a 
straightforward way \cite{Kassubek99,StaffordPRL99,StaffordEDP00}, but
do not alter our main conclusions.

\section{Quantum Transport}

Evaluating the transmission probabilities $\{T_n\}$ in the WKB approximation
for an axially-symmetric nanowire \cite{StaffordPRL97}, 
the conductance calculated from
Eq.\ (\ref{eq:landauer}) is shown in the upper-right panel
of Fig.\ \ref{fig:fandg}.
Plateaus in the conductance at integer multiples of $G_0$ are evident, with
some rounding of the steps due to tunneling.  Some integers are absent,
reflecting the degeneracies associated with axial symmetry \cite{Krans,Torres}.

\begin{figure} 
\vspace*{-2mm}
\resizebox{9cm}{!}{\includegraphics{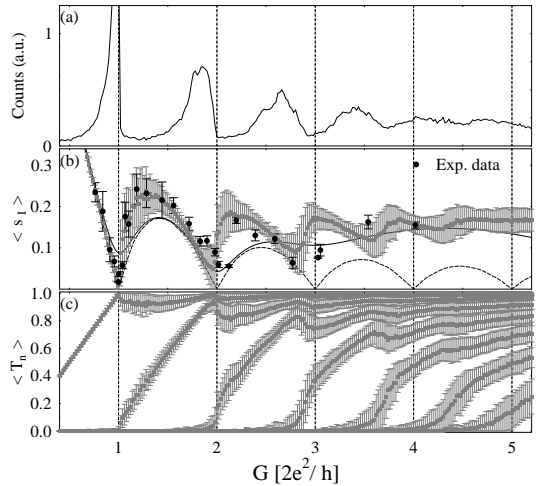}}
\vspace*{-2mm}
\caption{
(a) Calculated conductance histogram \cite{BuerkiPRB99}; (b)
calculated mean shot noise $\left<s_I\right>$ (grey squares \cite{BuerkiPRL99}),
together with experimental data from Ref.\ \cite{Brom99} (black circles);
(c) mean transmission probabilities $\left< T_n\right>$ \cite{BuerkiPRL99}.
The error bars indicate the
standard deviations of the numerical results over the ensemble
and the experimental errors, respectively.}
\label{fig:noise}
\end{figure}

Conductance steps of size $G_0$ were first observed in quantum point contacts
(QPCs)
fabricated in semiconductor heterostructures \cite{Datta} 
and are a rather universal phenomenon in metal nanowires 
[1-4],
even being found in contacts formed in liquid metals \cite{liquid}.
The precision of conductance
quantization in metal nanowires is poorer than that in semiconductor
QPCs due to their inherently rough structure on the scale of the
Fermi wavelength $\lambda_F$, which causes backscattering \cite{BuerkiPRB99},
and due to the imperfect hybridization of the atomic orbitals in the
contact, especially for multivalent atoms \cite{Scheer}.
For this reason, a statistical analysis of data for a large number of contacts
is often made \cite{Olesen,Krans,liquid,Brom99,Yanson}, resulting in a {\em
conductance histogram} [see Fig.\ \ref{fig:noise}(a)].

To model quantum transport in gold nanowires, 
where there are no ``missing integers'' in the conductance histogram
\cite{Olesen,liquid,Brom99},
geometries without axial symmetry were chosen, and weak disorder,
corresponding to a mean-free path $k_F \ell=270$, was included 
both in the nanowire and in the electrodes neighboring it \cite{BuerkiPRB99}.
The transmission probabilities were calculated by solving Schr\"odinger's 
equation using a recursive Green's function algorithm \cite{BuerkiPRB99}.
Averaging over different contact shapes and impurity configurations, we
obtained the histogram shown in Fig.\ \ref{fig:noise}(a), which is very
similar to typical experimental histograms for gold
\cite{Olesen,liquid,Brom99}.  The effect of disorder is twofold 
\cite{BuerkiPRB99}: the conductance
peaks are shifted downward due to backscattering, and the peaks are 
broadened due to universal conductance fluctuations, filtered by the nanowire.

Recently, additional information on quantum transport in metal
nanowires has been obtained from experiments on shot noise
\cite{Brom99}.
{\em Shot noise} is the term used to describe the
temporal fluctuations of electric current arising from the discreteness of
the electric charge $e$.  In 1918, Schottky showed
that if the arrival times of charge carriers are uncorrelated,
the shot-noise spectral power
$P_I=2eI$, where $I$ is the time-average current.  However,
in a quantum conductor with a finite number of transmitted
modes, the shot noise is suppressed below the Schottky value due to
anticorrelations induced by Fermi-Dirac statistics.
The suppression factor at zero temperature is given by \cite{Brom99}
\begin{equation}
s_I = \frac{P_I}{2eI} = \frac{\sum_n T_n(1-T_n)}{\sum_n T_n}.
\label{eq:noise}
\end{equation}
Fig.\ \ref{fig:noise}(b) shows the measured shot noise (solid circles
\cite{Brom99}) of gold nanowires as a function of their 
conductance.  The pronounced suppression of $s_I$ for wires with
conductances near integer multiples of $G_0$ reveals unambiguously
the quantized nature of the electronic transport.
We computed \cite{BuerkiPRL99}
the mean and standard deviation of
$s_I$ and $T_n$ as functions of $G$ (grey squares in Fig.\ \ref{fig:noise})
from the
numerical data used to generate the conductance histogram in 
Fig.\ \ref{fig:noise}(a).
The agreement of the experimental results for
particular contacts and the calculated distribution of $s_I$ shown
in Fig.\ \ref{fig:noise}(b) is extremely good: 67\% of the experimental
points lie within one standard deviation of $\langle s_I \rangle$
and 89\%  lie within two standard deviations.
It should be emphasized that no attempt has been made to fit the
shot-noise data; the numerical data of Ref.\ \cite{BuerkiPRB99}, where
the length of the contact and the strength of the disorder
were chosen to model experimental
conductance histograms for gold, have simply
been reanalyzed to calculate $\langle s_I \rangle$.
The 97\% suppression of shot noise for nanowires with a single quantum
of conductance (i.e., wires one atom thick) suggests that such wires could be
useful for low-temperature/low-noise applications, such as quantum computing.

\section{Metallic nanocohesion}

The cohesive force of the nanowire is $F=-\partial \Omega/\partial L$,
where $L$ is the length of the nanowire.  We assume that the volume per
atom is conserved under elongation (ideal plastic deformation), 
so that the deformation occurs at
constant volume (for alternative constraints, see Refs.\ 
\cite{StaffordPRL99,StaffordEDP00}).
While the conductance is determined by the transmission probabilities, Eqs.\
(\ref{eq:omega_def}) and (\ref{eq:dos_dmb}) indicate that the energetics
of a nanowire are determined by the scattering {\em phase shifts}.
Evaluating the phase shifts in the WKB approximation, performing the 
energy integral in Eq.\ (\ref{eq:omega_def}) at $T=0$, 
and taking the derivative with respect to elongation
\cite{StaffordPRL97}, one finds 
the force shown in the lower-right panel of Fig.\
\ref{fig:fandg}.  The correlations between the force and conductance 
are striking: as the wire is elongated and its diameter decreases, $|F|$ 
increases along a conductance plateau, but decreases sharply
when the conductance drops.  
Each transmitted mode acts like a delocalized metallic bond, which can
be stretched and broken.

The calculated force is remarkably similar,
both quantitatively and qualitatively,
to the measured force for gold nanowires, shown in the lower-left panel
of Fig.\ \ref{fig:fandg}.  Inserting the value $\varepsilon_F/\lambda_F
\simeq 1.7\mbox{nN}$ for gold, we see that both the overall scale of the 
force for a given value of the conductance and the heights of the last
two force oscillations are in quantitative agreement with the experimental
data.  One discrepancy is that the jumps in both force and conductance 
are less abrupt than in the experimental curves, possibly because we 
considered only geometries that change continuously with elongation.

In order to separate out the mesoscopic sawtooth structure in the force,
associated with the opening of individual conductance channels, from the
overall (macroscopic)
trend of the contact to become stronger as its diameter increases, it is
useful to perform a systematic semiclassical expansion
\cite{Gutzwiller,SemiclPhys} of the DOS,
$g(\varepsilon) = \bar{g}(\varepsilon) + \delta g(\varepsilon)$,
where $\bar{g}$ is a smooth average term, referred to as the
Weyl contribution, and $\delta g(\varepsilon)$ 
is an oscillatory term, whose average
is zero.  For the free electron model with Dirichlet boundary conditions,
the Weyl term is \cite{SemiclPhys}
\begin{equation}
\bar{g}(\varepsilon) =
\varepsilon^{-1}\left(\frac{k^3 V}{2\pi^2}
- \frac{k^2 A}{8\pi}
+ \frac{k \, C}{6\pi^2} \right),
\label{eq:dos_weyl}
\end{equation}
where $k=\sqrt{2m\varepsilon}/\hbar$,
$V$ is the volume of the wire, $A$ its
surface area, and $C$ the integrated mean curvature of its surface.
The oscillatory contribution $\delta g(\varepsilon)$ to the DOS
may be approximated as a Feynman sum
over classical periodic orbits \`a la Gutzwiller \cite{Gutzwiller,SemiclPhys}:
\begin{equation}
\delta g(\varepsilon) = 
\sum_\nu A_\nu 
\cos\left(\frac{S_\nu(\varepsilon)}{\hbar}
+\theta_\nu\right),
\label{eq:dos_gutzwiller}
\end{equation}
where $S_\nu$ is the classical action of a periodic orbit, $\theta_\nu$
is a phase shift determined by the singular points along the
classical trajectory, and $A_\nu$ is an amplitude
depending on the stability, symmetry, and period of the orbit.
Using $\bar{g}(\varepsilon)$ in Eq.\ (\ref{eq:omega_def}), one can derive
a Sharvin-like formula for the force
\begin{equation}
F=\bar{F} + \delta F, \qquad \bar{F}= -\frac{\varepsilon_F}{\lambda_F} 
\left(\frac{\pi k_F D}{16} - \frac{4}{9}\right).
\label{eq:f_sharvin}
\end{equation}
The first term in $\bar{F}$ is the {\em surface tension}.
The oscillatory mesoscopic correction $\delta F$
may be calculated with the aid of Eq.\ (\ref{eq:dos_gutzwiller}).
Under reasonable assumptions about the geometry, it can be shown 
\cite{StaffordPRL99} that the amplitude of the force oscillations is
{\em universal}:
\begin{equation}
\mbox{rms}\, \delta F = 0.58621 \, \varepsilon_F/\lambda_F.
\end{equation}

In more realistic models including electron-electron interactions
\cite{Kassubek99,StaffordPRL99,StaffordEDP00} and self-consistent
confining potentials \cite{Yannouleas98}, the surface tension is 
typically reduced compared to Eq.\ (\ref{eq:f_sharvin}), but the 
force oscillations are essentially the same as in the free-electron model.

\section{Stability of Nanowires}

\begin{figure} 
\vspace*{-5mm}
\resizebox{8.0cm}{!}{
\includegraphics*{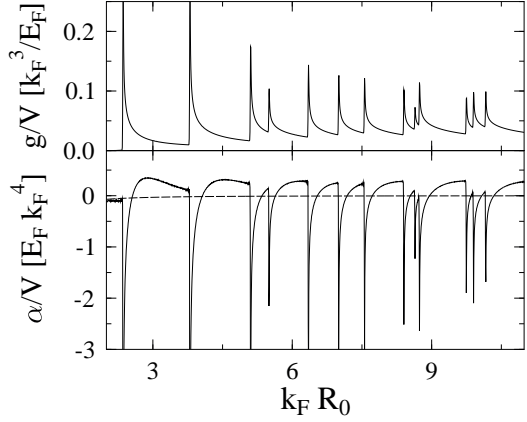}}
\vspace*{-5mm}
\caption{Density of states $g(\varepsilon_F)$ of a cylindrical wire
(upper diagram) and zero temperature
stability coefficient $\alpha$ (lower diagram) versus the
radius $R_0$ of the unperturbed wire.
The wavevector of the perturbation is $q R_0 =1$.
Dashed curve: Weyl contribution to $\alpha$.
}
\label{fig:alpha}
\end{figure}

A cylindrical body longer than its circumference is unstable to breakup
under surface tension \cite{Chandrasekhar}, a phenomenon
known as the {\em Rayleigh instability}.   
How then to explain the durability of long
gold nanowires [c.f.\ Fig.\ \ref{fig:nanobridge}(b)], the thinnest of
which have been shown \cite{Kondo00}
to be almost perfectly cylindrical in shape?
The key is the quantum corrections
\cite{Kassubek01} to the classical stability coefficients.

Only axially-symmetric deformations can lower the surface energy
of a cylindrical object, and thus lead to an instability
\cite{Chandrasekhar}.  Any such
deformation may be written as a Fourier series
\begin{equation}
R(z)=R_0 +\int_{-\infty}^\infty dq\, b(q) e^{iqz},
\end{equation}
where $R(z)$ is the radius of the wire at $z$, $R_0$ is the 
radius of the unperturbed cylinder, 
and $b(q)$ is a complex perturbation coefficient.
Using Eqs.\ (\ref{eq:dos_weyl}) and (\ref{eq:dos_gutzwiller}) in Eq.\ 
(\ref{eq:omega_def}), one obtains the following expansion \cite{Kassubek01}
\begin{equation}
\Omega[b]=\Omega[0]+\int_0^\infty\!\! dq\, \alpha(q) |b(q)|^2
+ {\cal{O}}(b^3),
\label{eq:omega_pert}
\end{equation}
where the stability coefficient $\alpha(q)$ depends implicitly on
$R_0$ and $T$.
The change in the grand canonical potential is of second order in $b$
and contributions from deformations with different $q$ decouple. If
$\alpha(q)$ is negative for any value of $q$, then $\Omega$
decreases under the deformation and the wire is unstable.

Let us first discuss the stability of a nanowire at zero temperature.
Fig.\ \ref{fig:alpha} shows the stability coefficient (lower diagram) and 
DOS (upper diagram) at the
classical stability threshold $qR_0=1$ as a function of $R_0$.  
For a straight wire, the transverse motion is quantized, and the 
DOS consists of a sequence of sharp peaks associated with the 
opening of each successive subband.
$\alpha$
has sharp negative peaks---indicating strong instabilities---at
the subband thresholds, where the density of states is sharply peaked.
Under surface tension and curvature energy alone
(dashed curve in Fig.\ \ref{fig:alpha}), the wire
would be slightly unstable at the critical
wavevector $q R_0=1$, since the curvature term is negative.
However, the quantum correction is positive in the
regions between the thresholds to open new subbands, {\em
thus stabilizing the wire}.  Since the oscillatory contribution to $\alpha$
is independent of $q$, we find that regions of stability persist
for {\em arbitrarilly long wavelength perturbations}, indicating that
an infinitely long cylindrical wire is a true metastable state if the radius
lies in one of the windows of stability.

\begin{figure} 
\vspace*{-2.1cm}
\resizebox{7.5cm}{!}{\includegraphics{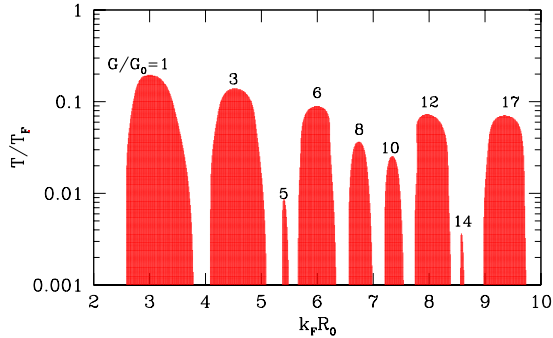}}
\vspace*{-7mm}
\caption{Stability of metal nanowires as a function of radius and 
temperature.  Shaded regions indicate stability
with respect to small perturbations; unshaded regions denote unstable
configurations.  Here $T_F$ is the Fermi temperature and $R_0$ is the 
mean radius of the wire.  The quantized conductance values of the 
stable wires are indicated.
}
\label{fig:stability}
\end{figure}

With these results, we can construct a stability diagram for metal
nanowires (see Fig.\ \ref{fig:stability}).  In the semiclassical approximation,
$d\alpha/dq > 0$ always, so the stability of the wire is determined by the
sign of $\alpha(q=0)$.  $\alpha(0)$ is a function of two dimensionless
parameters, $k_F R_0$ and $T/T_F$.  Regions where $\alpha(0)>0$ are 
shaded dark in Fig.\ \ref{fig:stability}, while regions where $\alpha(0)<0$ are 
unshaded.  The stable regions persist up to extremely high
temperatures for several quantized conductance values
(recall that $T_F > 10^4\mbox{K}$ for metals), 
indicating that {\em electron-shell effects
may stabilize nanowires even for temperatures well above the bulk melting
temperature}.
It is important to point out that if a more realistic value of the 
surface tension \cite{StaffordPRL99,StaffordEDP00,Yannouleas98} were
used, the stability boundaries would be pushed to 
{\em even higher temperatures}.
Thus the astounding stability properties shown in Fig.\ \ref{fig:stability}
are a very robust prediction of the jellium model.

\section{Conclusions}
Wires formed from chains of individual metal atoms 
have a number of properties which make them promising for
nanotechnology: They are very strong, able to support tensions up to
$\varepsilon_F/\lambda_F \approx 1\mbox{nN}$.  Contrary to naive expectations,
they are extremely stable, despite their large surface to volume ratio.
They are nearly-ideal one-dimensional conductors, and exhibit 
dramatically-reduced shot noise.  One potential application of metal nanowires 
is for integrated-circuit interconnects, due to their high conductance and
structural robustness.  The quantum suppression of shot noise in nanowires
may also make them useful for low-temperature/low-noise
applications, such as quantum computing.
\\
\\
\textbf{\emph{Acknowledgements}}\indent 
I am indebted to 
J\'er\^ome B\"urki, Frank Kassubek, and Chang-hua Zhang,
the students and postdocs without whom much of this research would have been
impossible.  I also wish to thank Dionys Baeriswyl, Raymond Goldstein, 
Hermann Grabert, and Xenophon Zotos for their valuable contributions.
This research was supported by NSF grant DMR0072703 and
by an award from Research Corporation.

\end{document}